\documentclass[12pt]{revtex4}
\usepackage[dvips]{graphicx}
\usepackage{amssymb}
\usepackage{color}

 \newcommand{\g}[1]{\mbox{\boldmath $#1$}}

 \newcommand{\lp}{\left(}
\newcommand{\rp}{\right)} 
\newcommand{\mygtrsim}{\mathrel{\mbox{\raisebox{-1mm}{$\stackrel{>}{\sim}$}}}}

\begin{document}
\begin{center}
\title{Electromagnetic waves destabilized by runaway electrons in  near-critical electric fields}

\author{A. K\'om\'ar$^{1}$, G.I. Pokol$^{1}$, T. F\"ul\"op$^{2}$}

\affiliation{
\small $^{1}$ Department of Nuclear Techniques, Budapest University of Technology and Economics, Association EURATOM, H-1111 Budapest, Hungary\\
\small  $^{2}$ Department of Applied Physics, Nuclear Engineering, Chalmers University of Technology and Euratom-VR Association, G\"oteborg, Sweden\\
}

\maketitle

\end{center}
\begin{abstract}
  Runaway electron distributions are strongly anisotropic in velocity
  space. This anisotropy is a source of free energy that may
  destabilize electromagnetic waves through a
  resonant interaction between the waves and the energetic
  electrons. In this work we investigate the high-frequency
  electromagnetic waves that are destabilized by runaway electron
  beams when the electric field is close to the critical field for
  runaway acceleration. Using a runaway electron distribution
  appropriate for the near-critical case we calculate the linear
  instability growth rate of these waves and conclude that the
  obliquely propagating whistler waves are most unstable. We show that
  the frequencies, wave numbers and propagation angles of the most
  unstable waves depend strongly on the magnetic field. Taking into
  account collisional and convective damping of the waves, we
  determine the number density of runaways that is required to
  destabilize the waves and show its parametric dependences.
\end{abstract}
\maketitle

\section{Introduction}
Relativistic runaway electron populations have been frequently
observed in various plasmas, e.g. large tokamak disruptions \cite{td},
electric discharges associated with thunderstorms \cite{tavani} and solar flares
\cite{moghaddam}.  Runaway electrons are produced when the electric
field is larger than a certain critical field ($E_c$), and the
accelerating force overwhelms the friction for high energy electrons.
The anisotropy of the runaway electron distribution can lead to
destabilization of electromagnetic waves through
wave-particle resonant interaction.  Several studies have shown that
the velocity anisotropy excites electromagnetic waves
mainly through the anomalous Doppler resonance
\cite{fulop,pokol}. Once the instability is triggered the
distribution is isotropized due to pitch-angle scattering. Previous
work \cite{pokol,fulop,fulop1} has considered whistler wave
instability driven by an anisotropic electron distribution as a
possible cause for the observed magnetic field threshold for runaway
generation in large tokamaks \cite{gill,jt60}. These calculations
relied on a distribution function that was based on an approximate
solution of the kinetic equation in the case when the electric field
is well above the critical field, $\alpha\gg 1$, where
\begin{equation}
\alpha=\frac{E}{E_c}=\frac{4 \pi \epsilon_0^2m_e c^2}{n_e e^3 \ln{\Lambda}}E,
\end{equation}
where $n_e$ is the thermal electron density, $m_e$ is the electron
rest mass, $e$ is the electron charge, $\ln{\Lambda}$ is the Coulomb
logarithm, $\epsilon_0$ is the dielectric constant and $c$ is the
speed of light. Also, in many studies, the runaway electrons were
assumed to be ultra-relativistic (velocities within $5
  \%$ to the speed of light) and simplified resonance conditions were
used to describe the wave-particle interaction. However, the electric
field is not always much larger than the critical field and the
velocity of the electrons is often not that close to the speed of
light. An example of this is the observations of
  superthermal electron populations in the T-10 tokamak during
  magnetic reconnection events, when the electric field was
  transiently larger than the critical field during the reconnection
  (for about $0.1\;\rm ms$) but then it dropped to values near or even
  below the critical field \cite{savrukhin}. Recent work
\cite{riemann} has shown that even in disruptions the electric field
in the core region of the plasma is only slightly above the critical
electric field, $\alpha {\mygtrsim} 1$.

The purpose of this work is to determine what waves could be
destabilized by runaway beams in a near-critical
field. Investigating the lowest relevant limit of the
  electric field when runaway production occurs is a step toward
  generalizing the analysis of the runaway electron driven
  instabilities to lower electric fields. This way we can gain
  confidence that the analysis of the wave-particle interaction yields
  valid results in both the high electric field and the near-critical
  limit, before proceeding to the numerical analysis of the
  interaction for electric fields in between.

In the present work we use the runaway distribution derived in
Reference~\cite{sandquist}, appropriate for a near-critical field, to
calculate the instability growth rate of these waves and determine the
frequencies and wave numbers of the most unstable waves for various
parameters. We use a general resonance condition, without the
ultra-relativistic assumption, so the model can be applied also for
electrons with lower energies. We show that the whistler branch is
destabilized via the anomalous Doppler and Cherenkov resonances.
Increasing magnetic field leads to increasing wave number and
frequency while decreasing propagation angle for the most unstable
wave. The observation of these waves could help to determine the
origin and evolution of the energetic electrons. If the waves grow to
significant amplitude they may contribute to efficient transport of
particles out from the plasma.

The remainder of the paper is organized as follows.  In
Sec.~\ref{sec:dispersion}, the wave dispersion equation is presented,
together with a perturbative approximation of the instability growth
rate.  In Sec.~\ref{sec:distr} the runaway electron distribution in a
near-critical electric field is analyzed and the runaway contribution
to the susceptibilities is calculated. In Sec.~\ref{sec:growth} the
instability growth rate of the high-frequency electromagnetic waves
driven by runaways is calculated and the parameters of the most
unstable wave are determined. Here, we also show the stability
thresholds of the waves and study their parametric
dependences. Finally, the results are summarized and discussed in
Sec.~\ref{sec:conclusions}.

\section{Dispersion relation}
\label{sec:dispersion}
The dispersion relation of high frequency electromagnetic waves is
given by \cite{stix}
\begin{equation}
\lp \epsilon_{11}-k_\parallel^2 c^2/\omega^2\rp\lp\epsilon_{22}-k^2c^2/\omega^2\rp+\epsilon_{12}^2=0,
\label{eq:2x2dispersion}
\end{equation}
where $\omega$ is the wave frequency, $\textbf{k}$ the wave number and
$\mbox{\boldmath${\epsilon}$}$ the dielectric tensor of the
plasma. Equation (\ref{eq:2x2dispersion}) follows from the
wave-equation, with the approximation $\epsilon _{33} \gg n^2
\cos{\theta} \sin{\theta}$, where $\displaystyle \textbf{n} =
\textbf{k} c/\omega$ is a dimensionless vector with
  the magnitude of the refractive index,
$\cos{\theta}=k_\parallel/k$, $\theta$ is the pitch
  angle.  The subscripts $_\|$ and $_\perp$ denote the parallel and
perpendicular directions with respect to the magnetic field. The
dielectric tensor is
\begin{equation}
\mbox{\boldmath${\epsilon}$} = \textbf{1} + \mbox{\boldmath${\chi}$} ^i + \mbox{\boldmath${\chi}$} ^e + \mbox{\boldmath${\chi}$} ^r ,
\label{eq:dielectric0}
\end{equation}
where $\mbox{\boldmath${\chi}$} ^s$ is the susceptibility of plasma
species $s$, where $i$ denotes the ion, $e$ the thermal electron and
$r$ the runaway electron populations, $\textbf{1}$ is the dyadic
unit. As the contribution of the runaway population is expected to be
small, we consider the dispersion of high frequency electromagnetic
waves without the runaway term and use the cold plasma approximation
\cite{stix} for the ion and electron populations. The
contribution of runaway electrons is added later as a perturbation,
which is justified in the present case since the
  runaway electron density is much smaller than the thermal electron
  and ion density.
\subsection{Electron-whistler wave}
\begin{figure}[htbp]
  \begin{center}
	\includegraphics[width=0.9\textwidth]{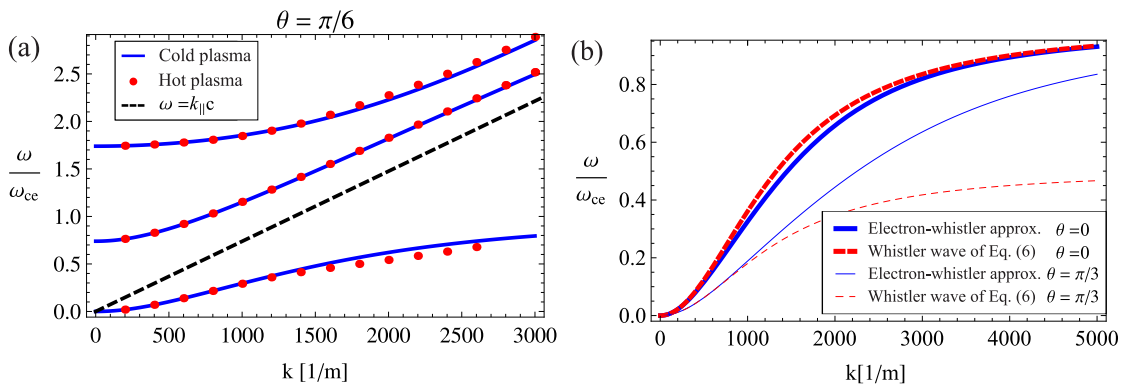}
	 \caption{(a) Solution
      of the analytical approximation of the dispersion relation from
      Equation~(\ref{fomega}) (solid) together with the numerical solution
      using the hot plasma susceptibilities, for plasma temperature
      $T=20\;\rm keV$, density $n_e = n_i = 5 \cdot 10^{19}\;\rm
      m^{-3}$, magnetic field $B = 2\;\rm T$ and propagation angle
      $\theta=\pi/6$. Dashed line shows $\omega=k_\parallel c$. For
      the electron-whistler wave $\omega<k_\parallel
      c$. (b) The lowest frequency solution of the analytical
      approximation of the dispersion relation Equation~(\ref{fomega})
      (blue solid) together with the whistler approximation from
      Equation~(\ref{ww}) (red dashed), for propagation angles $\theta= 0$ (thick
      lines) and $\theta=\pi/3$ (thin lines).  }
\label{fig:disp}
\end{center}\end{figure}

For the frequency range $\omega_{ce}{\sqrt{m_e/m_i}} \ll\omega$, the
background ion and electron contributions to $\g \epsilon$ are
\cite{stix}
\begin{equation}
\begin{array}{ccc}
\displaystyle\epsilon_{11}^{e+i}=\epsilon_{22}^{e+i}=1-\frac{\omega_{pe}^2}{\omega^2-\omega_{ce}^2}
&\mbox{and} 
&\displaystyle\epsilon_{12}^{e+i}=-i \frac{\omega_{pe}^2\omega_{ce}}{\omega(\omega^2-\omega_{ce}^2)}.
\end{array}
\label{eq:chi-electron}
\end{equation}
Here  $\omega_{pe}$ and $\omega_{ce}$ are the electron plasma and
cyclotron frequencies, respectively.
 Without runaways, the dispersion relation can be written as
\begin{eqnarray}
\mathcal{E}(\omega)\equiv \omega^6-\omega^4[2 \omega_{pe}^2+\omega_{ce}^2+(k^2+k_\parallel^2)c^2]\nonumber \\+\omega^2[\omega_{pe}^4+(k^2+k_\parallel^2)c^2(\omega_{pe}^2+\omega_{ce}^2)+k^2 k_\parallel^2 c^4]-k^2 k_\parallel^2 c^4 \omega_{ce}^2=0.
\label{fomega}
\end{eqnarray}
Equation (\ref{fomega}) has three solutions for $\omega^2$ and these
can be determined analytically, although their closed form expressions
are very complicated. One of the solutions satisfies $\omega <
k_{\parallel} c $ for all wave numbers $k$ and propagation angles
$\theta$ and will be called `electron-whistler' wave, because in
certain limits, as we will show, its dispersion characteristics are
the same as the whistler wave's.  For the two other solutions
$\omega > k_{\parallel} c$ is satisfied.  For typical
experimental parameters, the solution of the analytical dispersion
relation (\ref{fomega}) has excellent agreement with the numerical
solution of the full dispersion relation using the hot plasma
susceptibilities for both ions and electrons from Reference~\cite{stix}
instead of Equation~(\ref{eq:chi-electron}).  Figure \ref{fig:disp}a shows
the three solutions of Equation~(\ref{fomega}) together with the solution
of the numerical dispersion relation. The solution for the wave
frequency is plotted as function of wave number for propagation angle
$\theta=\pi/6$. The wave dispersions in Figure \ref{fig:disp} are
calculated for $T=20\;\rm keV$. The agreement is even better at lower
temperatures.

The whistler approximation is usually defined by \cite{ww}
\begin{equation}
\frac{k^2c^2}{\omega^2}\left(\frac{\omega_{ce}}{\omega} \cos{\theta}-1\right)=\frac{\omega_{pe}^2}{\omega^2}.
\label{ww}
\end{equation}
To show the whistler character of the lowest frequency solution of
$\mathcal{E}(\omega)=0$, we plot it together with the solution of
Equation~(\ref{ww}) as functions of $k$ for $\theta=0$ and
$\theta=\pi/3$, see Figure~\ref{fig:disp}b. For $\theta=0$ there is
very good agreement between the two solutions. For $\theta=\pi/3$ and
wave numbers up to 1000 $\rm m^{-1}$ the two solutions overlap, but
for higher wave numbers they deviate. This difference
  is due to the approximation $\epsilon _{33} \gg n^2 \cos{\theta}
  \sin{\theta}$ used when deriving Equation (\ref{eq:2x2dispersion}),
  while when deriving the whistler wave dispersion given in Equation
  (\ref{ww}) in Reference \cite{ww} no such approximation was used.
  By investigating the validity of the (\ref{fomega}) dispersion we
  concluded that it yields valid results compared to the general
  dispersion relation for magnetic fields up to $3 \;\rm T$. In the following we will therefore limit our analysis to $B<3 \;\rm T$.

Including runaways, Equation~(\ref{fomega}) can be written as
\begin{eqnarray}
\hspace{-1cm} \mathcal{E}(\omega)=\omega^4(\omega^2-\omega_{ce}^2)\left[\chi_{11}^{r}\left(\frac{k^2 c^2}{\omega^2}-\epsilon_{22}^{e+i}\right)+\chi_{22}^{r}\left(\frac{k_\parallel^2 c^2}{\omega^2}-\epsilon_{11}^{e+i}\right)-2 \epsilon_{12}^{e+i}\chi_{12}^r\right],
\label{eq:disp1}
\end{eqnarray}
where $\chi_{ij}^{r}$ denotes the runaway contribution to the susceptibility tensor. The linear growth rate of a small perturbation of the wave frequency $\omega=\omega_0+\delta\omega$, is $\gamma_{i}= \Im\delta\omega$ and is given by
\begin{equation}
\frac{\gamma_{i}^e}{\omega_0}=\Im \frac{\omega_0^2(\omega_0^2-\omega_{ce}^2)\left[\chi_{11}^{r}\left(\frac{k^2 c^2}{\omega^2}-\epsilon_{22}^{0}\right)+\chi_{22}^{r}\left(\frac{k_\parallel^2 c^2}{\omega^2}-\epsilon_{11}^{0}\right)-2 \epsilon_{12}^{0}\chi_{12}^r\right]}{2\left\{3\omega_0^4-2 \omega_0^2[2\omega_{pe}^2+\omega_{ce}^2+(k^2+k_\parallel^2)c^2]+\omega_{pe}^4+(k^2+k_\parallel^2)c^2(\omega_{pe}^2+\omega_{ce}^2)+k^2k_\parallel^2 c^4\right\}},
\label{eq:growth-electron}
\end{equation} 
where $\Im$ denotes the imaginary part and  $\epsilon_{ij}^0$ are the cold plasma dielectric tensor elements {defined by Equation~(\ref{eq:chi-electron})} evaluated at the unperturbed wave frequency: $\epsilon_{ij}^0=\epsilon_{ij}^{e+i}(\omega=\omega_0)$.

\subsection{Magnetosonic-whistler}
To evaluate the wave-particle interaction in the lower frequency
region we analyze the dispersion relation in the frequency range
$\omega_{ci}\ll\omega\ll\omega_{ce}$. Interaction between these waves
and strongly relativistic runaways has been studied before
\cite{pokol,fulop}. However, in a near-critical field, it is more
likely that the runaways are mildly relativistic, and as both the
distribution function and the resonance condition is different, the
analysis in previous work has to be generalized. In this frequency range the
contributions to the dielectric tensor elements 
are
\begin{eqnarray}
\displaystyle\epsilon_{11}^{e+i}=1-\frac{\omega_{pi}^2}{\omega^2}+\frac{\omega_{pe}^2}{\omega_{ce}^2}\nonumber\\
\displaystyle\epsilon_{22}^{e+i}=1-\frac{\omega_{pi}^2}{\omega^2}+\frac{\omega_{pi}^2}{\omega_{ci}\omega_{ce}}\label{eq:chi-magnetosonic}\\
\displaystyle\epsilon_{12}^{e+i}=i \frac{\omega_{pi}^2}{\omega_{ci}\omega},
\nonumber
\end{eqnarray}
where $\omega_{pi}$ and $\omega_{ci}$ are the ion plasma and cyclotron
frequencies, respectively. Substituting these into Equation (\ref{eq:2x2dispersion}) leads to the following dispersion
relation
\begin{eqnarray}
\hspace{-2cm} k^2v_A^2\left( 1+\frac{k_\parallel^2v_A^2}{\omega_{ci}^2} +\frac{k_\parallel^2}{k^2}\right)-\omega^2\left( 1+ \frac{(k^2+k_\parallel^2-2 \omega^2/c^2) v_A^2}{\omega_{ci}\omega_{ce}}+\frac{ (k_\parallel^2+k^2) v_A^2}{\omega_{pi}^2}-\frac{v_A^2}{c^2}\frac{\omega^2}{\omega_{pi}^2}\right)\nonumber\\\equiv M(\omega) =0,
\label{momega}
\end{eqnarray}
where $v_A=c \omega_{ci}/\omega_{pi}$ is the Alfv\'en speed.  In  Reference~\cite{fulop} a
simplified version of this dispersion relation
\begin{equation}
 k^2v_A^2\left( 1+\frac{k_\parallel^2v_A^2}{\omega_{ci}^2} +\frac{k_\parallel^2}{k^2}\right)-\omega^2\left( 1+ \frac{(k^2+k_\parallel^2) v_A^2}{\omega_{ci}\omega_{ce}}\right)\equiv M_s(\omega)=0,
\label{eq:magn-whistler-full}
\end{equation}
valid in the limit $\omega^2\ll k_\parallel^2 c^2$, was used to study
destabilization of waves by an avalanching runaway electron
distribution.  The wave determined by Equation~({\ref{momega}})
can be identified as the generalized magnetosonic-whistler
wave{, as its simplified limit,
  Equation~(\ref{eq:magn-whistler-full}),} for quasi-perpendicular
propagation $|k|\gg |k_\parallel|$ and $k^2 c^2\ll \omega_{pe}^2$
\begin{equation}
k^2v_A^2\left( 1+\frac{k_\parallel^2c^2}{\omega_{pi}^2} \right)-\omega^2=0.
\label{eq:magn-whistler-approx}
\end{equation}
has been previously identified as the magnetosonic-whistler wave \cite{fulop}.

\begin{figure}[htbp]
  \begin{center}
	\includegraphics[width=0.9\textwidth]{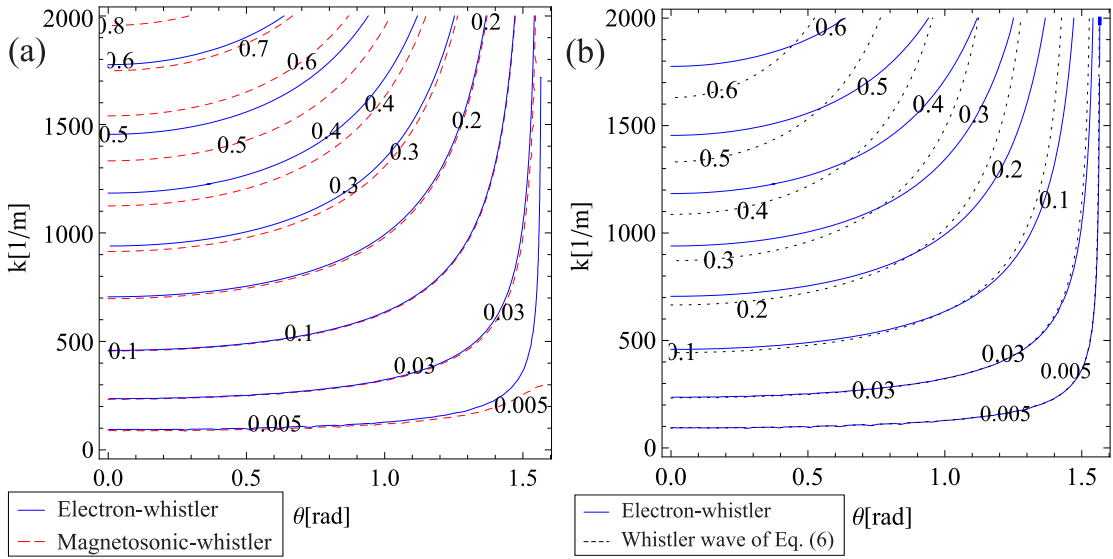}
    \caption{(a) Comparison of the lowest frequency solution of
      Equation~(\ref{fomega}) (blue solid) with the magnetosonic-whistler
      wave {of Equation~(\ref{momega})}  (red dashed). The parameters are the same as
      in Figure~\ref{fig:disp}.  (b) Contour plot of the lowest frequency
      solution of {Equation~(\ref{fomega})} and the solution of Equation~(\ref{ww}). The values plotted are $\omega/\omega_{ce}$ on both figures.}
\label{fig:whistlers}
\end{center}
\end{figure}
Figure \ref{fig:whistlers} shows contour plots of the electron-whistler wave together with the magnetosonic-whistler wave
(Figure~\ref{fig:whistlers}a) and the electron-whistler wave together
with the whistler approximation from Equation~(\ref{ww})
(Figure~\ref{fig:whistlers}b). The electron-whistler and magnetosonic-whistler waves approximately overlap with the whistler approximation
for low $k$ (any propagation angle) and quasi-perpendicular
propagation (any $k$). In the rest of the $k$-$\theta$-space the
electron-whistler and magnetosonic-whistler waves have different
dispersion characteristics, as the magnetosonic-whistler
approximation is not valid in the region of very high $k$ numbers
because its frequency is assumed to be $\omega\ll \omega_{ce}$.

Including runaways, {Equation~(\ref{momega})}  can be written as
$$
M(\omega)=\frac{\omega^2\omega_{ci}^2}{\omega_{pi}^2}\left[\chi_{11}^r\left(1-\frac{\omega^2}{\omega_{ci}\omega_{ce}}+\frac{k^2 v_A^2}{\omega_{ci}^2}-\frac{\omega^2}{\omega_{pi}^2}\right)+\chi_{22}^r\left(1-\frac{\omega^2}{\omega_{ci}\omega_{ce}}+\frac{k_\parallel^2 v_A^2}{\omega_{ci}^2}-\frac{\omega^2}{\omega_{pi}^2}\right)-2 i \frac{\omega}{\omega_{ci}}\chi_{12}^r\right]
$$
and the linear growth rate of a perturbation of the wave frequency is
\begin{equation}
\frac{\gamma_i^m}{\omega_0}=-\Im{\frac{\omega_{ci}^2\left[\chi_{11}^r\left(1-\frac{\omega_0^2}{\omega_{ci}\omega_{ce}}+\frac{k^2 v_A^2}{\omega_{ci}^2}-\frac{\omega_0^2}{\omega_{pi}^2}\right)+\chi_{22}^r\left(1-\frac{\omega_0^2}{\omega_{ci}\omega_{ce}}+\frac{k_\parallel^2 v_A^2}{\omega_{ci}^2}-\frac{\omega_0^2}{\omega_{pi}^2}\right)-2 i \frac{\omega_0}{\omega_{ci}}\chi_{12}^r\right]}
{2\left[ \omega_{pi}^2+ (k^2+k_\parallel^2-4 \omega_0^2/c^2) c^2(\omega_{ci}/\omega_{ce})+(k_\parallel^2+k^2-2 \omega_0^2/c^2) v_A^2\right]}}.
\label{eq: growth-magneto}
\end{equation}
In the following section we will calculate the runaway contribution to
the susceptibilities which will allow us to evaluate the linear growth
rate of the wave. 

\section{Runaway contribution}
\label{sec:distr}

The susceptibility due to the runaway electron population is given by \cite{stix}:
\begin{equation}
\g \chi^r = \frac{\omega _{pr} ^2}{\omega \omega _{cr}} \sum{\int ^{\infty} _0 2 \pi p_\perp dp_\perp \int ^{\infty} _{- \infty} dp_\parallel \frac{\Omega_e \textbf{S}_m}{\omega - k_\parallel v_\parallel - m \Omega_e}},
\label{eq:runaway-susceptibility}
\end{equation}
where
\begin{center}
$\displaystyle \textbf{S} _m = \left[ \begin{array}{cc} \frac{m^2 J_m ^2}{z^2} p_\perp U & \it{i} m \frac{J_m J'_m}{z} p_\perp U \\ - \it{i} m \frac{J_m J'_m}{z} p_\perp U & (J'_m) ^2 p_\perp U \end{array} \right],$
\vspace{0.5 cm}
$\displaystyle U = \frac{\partial f_r}{\partial p_\perp} + \frac{k_\parallel}{\omega} \left( v_\perp \frac{\partial f_r}{\partial p_\parallel} - v_\parallel \frac{\partial f_r}{\partial p_\perp} \right),$
\end{center}
$\Omega_e = \omega _{ce}/\gamma$ is the relativistic cyclotron frequency of the electrons, $J_m (z)$ is the Bessel function of the first kind, {$J'_m (z) = d J_m / dz$,} $\displaystyle z = k_\perp v_\perp /\Omega_e = k_\perp c p_\perp /\omega _{ce}$, $ p = \gamma v/c$ is the normalized relativistic momentum, $ \gamma = \sqrt{1 + p^2}$ is the relativistic factor, $f_r = f/n_r$ is the normalized runaway distribution and $m$ is the order of resonance.
The general (and implicit)
condition for the resonant momentum is
\begin{equation}
\label{eq:pres_implicit}
\displaystyle p_{\parallel} = \frac{\omega _0 \gamma - m \omega _{ce}}{k_{\parallel} c}.
\end{equation}
If the distribution function is known, the resonance condition allows
the integral in (\ref{eq:runaway-susceptibility}) to be evaluated
using the Landau prescription.
\subsection{Distribution of the runaway electrons}
\begin{figure}[htbp]
\begin{center}
\includegraphics[width=0.5\textwidth]{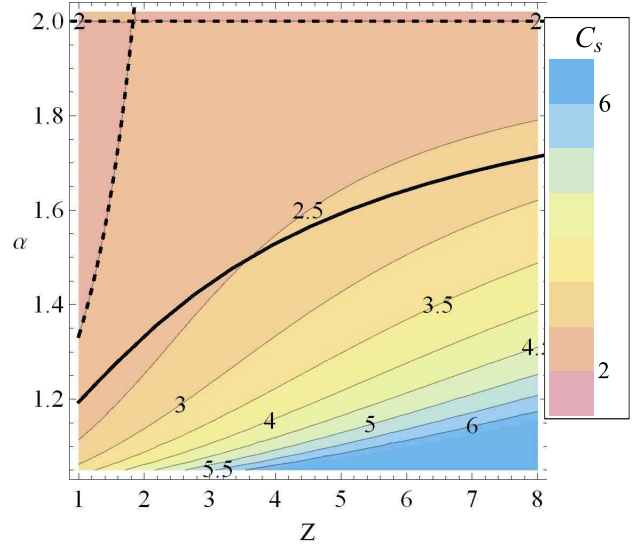}
\caption{$C_s$ as function of $\alpha$ and $Z$. The distribution
  function is valid in the region $2<C_s<1+\alpha$. Solid black line
  shows $C_s=1+\alpha$ and dashed black line is $C_s=2$. The region
  between the solid and dashed lines gives the combinations of
  $\alpha$ and $Z$ for which the condition $2<C_s<1+\alpha$ is
  fulfilled.}
\label{fig: alphaC_Z}
\end{center}
\end{figure}

To calculate the runaway susceptibilities, the runaway distribution
given in Equation~(83) of Reference~\cite{sandquist} is used for the
near-critical $\alpha\mygtrsim 1$ case
\begin{equation}
\hspace{-1cm}  f_r(p_\parallel, p_\perp)=\frac{A}{p_\parallel^{(C_s-2)/(\alpha-1)}}\exp{\left(-\frac{(\alpha+1)p_\perp^2}{2(1+Z) p_\parallel}\right)} \mbox{}_1F_1\left(1-\frac{C_s}{\alpha+1}, 1; \frac{(\alpha+1)p_\perp^2}{2 (1+Z) p_\parallel}\right),
\label{eq:near-distribution}
\end{equation}
where 
\begin{equation}
  C_s=\alpha-\frac{(1+Z)}{4}(\alpha -2) \sqrt{\frac{\alpha}{\alpha-1}},
\end{equation}
$Z$ is the effective ion charge and $\mbox{}_1F_1$ is the confluent
hypergeometric (Kummer) function. The distribution function given
above was obtained by matching asymptotic expansions in five separate
regions in momentum space. The calculation is similar to the one
presented by Connor and Hastie \cite{connorhastie} of runaway electron
generation, but it is valid for near-critical electric field. Note,
that to have a positive distribution function, the first argument of
$\mbox{}_1F_1$ should be positive, leading to the condition
$1>C_s/(\alpha+1)$. Furthermore, the condition $f_r\rightarrow 0$ as
$p_\parallel\rightarrow \infty$ requires that $C_s>2$. This gives a
region in the $\alpha$-$Z$ space where
Equation~(\ref{eq:near-distribution}) is valid.  The
parameter $C_s$ as function of $\alpha$ and $Z$ is plotted on
Figure~\ref{fig: alphaC_Z}.  The region between the solid and dashed
lines gives the combinations of $\alpha$ and $Z$ for which the
condition $2<C_s<1+\alpha$ is fulfilled.  This gives a restriction on
the effective charge number, since if $\alpha\simeq 1$, the charge
number can only be slightly more than unity. In tokamak plasmas $Z$
seldom exceeds values of about $3$.  In the following we will
only consider combinations of $\alpha$ and $Z$ such that
$2<C_s<1+\alpha$. One such combination is $\alpha=1.3$ and $Z=1$ and
this, together with the parameters $n_e = 5 \cdot 10^{19}\;\rm
m^{-3}$, $B=2\;\rm T$, are the baseline parameters of our study, and
will be used in the rest of the paper unless otherwise is stated.
Note, that although $C_s$ includes a term proportional
  to $1/\sqrt{\alpha-1}$, its value varies very little in the
  parameter space where the distribution function is valid, as
  Figure~\ref{fig: alphaC_Z} shows $C_s$ is between 2.5 and 3 in the
  region of interest, irrespective of the exact value of $\alpha$ and
  $Z$. Therefore the distribution function is not very sensitive to
  these values. 

\begin{figure}[htbp]
\begin{center}
\includegraphics[width=0.6\textwidth]{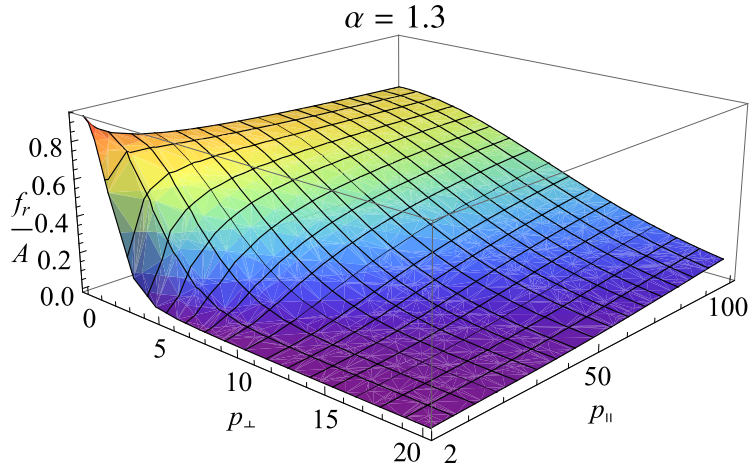}
\caption{Normalized runaway electron distribution function in near-critical
  field, $f_r/A$ plotted with respect to the parallel and perpendicular momentum normalized to $m_e c$, for $Z=1$ and $\alpha=1.3$.}
\label{fig:distr}
\end{center}
\end{figure}

Equation (\ref{eq:near-distribution}) is valid for all $p>p_c$ in the
case of near-critical electric field.  Note, that the integral of
Equation~(\ref{eq:near-distribution}) function in the whole momentum space
is divergent. This is because the electric field continuously
accelerates electrons and more and more electrons will run away.  In
spite of the continuous acceleration, the distribution is in
quasi-steady state, as the water leaking out of an unplugged bath tub
\cite{helander}. However, as the existence of the electric field is
finite in time, there is a maximum number of runaways and there is a
maximum energy which runaway electrons can reach in reality. In the
expressions for the runaway susceptibilities we use a normalized
distribution function $\int f_r d^3p=1$. The normalization constant
$A$ in Equation~(\ref{eq:near-distribution}) is obtained from
\begin{equation}
\int _0 ^\infty{ d p_\perp 2 \pi p_\perp \int _{p_c} ^{p_{\mathrm{max}}}{d p_\parallel \, f_r ( p_\parallel , p_\perp )}}=1,
\end{equation}
where $p_{\mathrm{max}}$ is the normalized momentum corresponding to the
maximum energy. This integral can be easily solved numerically if
$p_{\mathrm{max}}$ is known. The value of $p_{\mathrm{max}}$ depends on the exact value and time evolution of the accelerating field. In this paper we approximated the maximum energy as $2.6 \;\rm MeV$, corresponding to $p_{\mathrm{max}}=5$. A typical value of the perpendicular momentum can be determined from
  the runaway distribution function. In the case of $p_{\parallel max}
  = 5$, this is $p_\perp \approx 3$. This value corresponds to
  $E_\perp = 1.6$ MeV. 
Figure \ref{fig:distr} shows
Equation~(\ref{eq:near-distribution}) for 
$Z=1$ and $\alpha=1.3$. For the baseline parameters of our study this corresponds to the electric field of $0.06 \;\rm V/m$, while the critical field is $0.046 \;\rm V/m$.

It is instructive to compare the distribution in
Equation~(\ref{eq:near-distribution}) with the distribution derived for the
case of secondary runaway generation
\cite{pokol,fulop,fulop1} \begin{equation} f_r^{\mathrm{disr}} (p_\parallel, p_\perp)
  = \frac{a}{2 \pi c_Z p_{\parallel}} \exp{\left( \frac{-
        p_{\parallel}}{c_Z} - \frac{a p_{\perp} ^2}{2 p_{\parallel}}
    \right)},
\label{eq:high-ditribution}
\end{equation}
where $a=(\alpha-1)/(Z+1)$ and $c_Z = \sqrt{3 (Z+5)/\pi}
\ln{\Lambda}$. The  avalanche distribution is based on the
solution of the kinetic equation for relativistic electrons in the
limit of $\alpha\gg 1$ using the Rosenbluth-Putvinski runaway
growth rate \cite{rosput}
$$
\frac{dn_r}{dt}=\frac{n_r(\alpha-1)}{c_Z \tau}
$$
as boundary condition.  Here $\tau$ is the collision time for
relativistic electrons. This means that the runaway density grows
exponentially as $n_r=n_{r0} \exp{[(\alpha-1)t /(\tau c_Z)]}$, where
$n_{r0}$ is the seed produced by primary generation.  Note that for
the avalanching distribution $2 \pi \int p_\perp dp_\perp dp_\parallel
f_r=1$, independent of the maximum momentum. The distribution
(\ref{eq:high-ditribution}) is valid if secondary generation of
runaways is dominant, as expected to be the case in large tokamak
disruptions. In contrast, the distribution
(\ref{eq:near-distribution}) is valid when primary runaway production
is the main source of the superthermal electron population.

\begin{figure}[htbp]
\begin{center}
\includegraphics[width=0.9\textwidth]{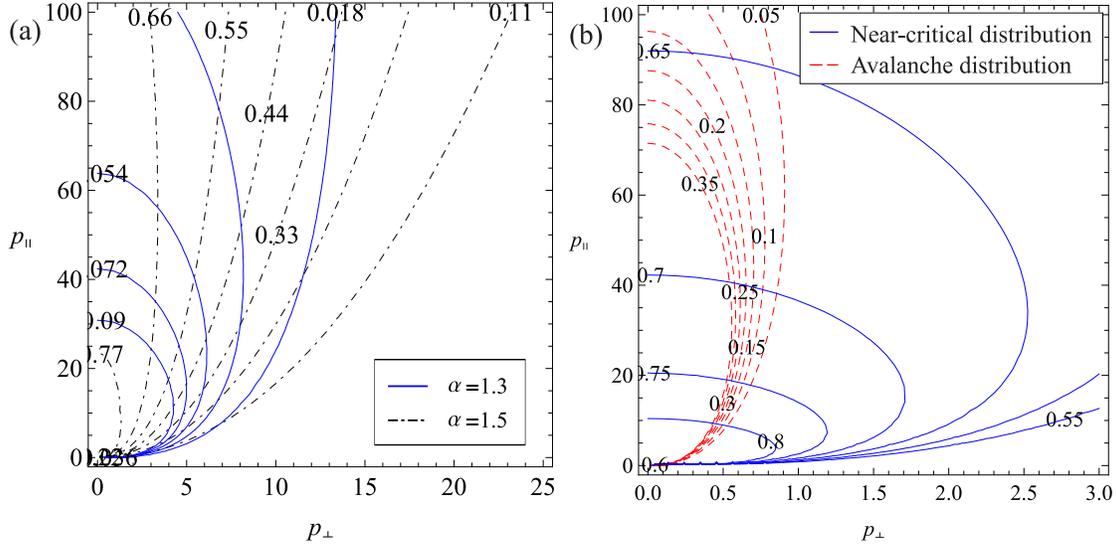}
\caption{(a) Contour plot of the distribution function,
  $f_r/A$ for $\alpha=1.3$ (solid, corresponding to $E=0.06 \;\rm V/m$) and $\alpha=1.5$
  (dashed, $E=0.069 \;\rm V/m$). The effective charge is $Z = 1.5$.  (b) Comparison between
  the near-critical, $f_r/A$ (blue solid) and
  avalanche, $100 f_r^{\mathrm{disr}}$ (red dashed) distribution functions. For the near-critical
  distribution we used $Z=1$ and $\alpha=1.3$. For
  the avalanche distribution we used $\ln{\Lambda}=18$, $Z=1$ and
  $E=40\;\rm V/m$ (corresponding to $\alpha=865$).  }
\label{fig:distr_critfield}
\end{center}
\end{figure}

\begin{figure}[htbp]
\begin{center}
\includegraphics[width=0.91\textwidth]{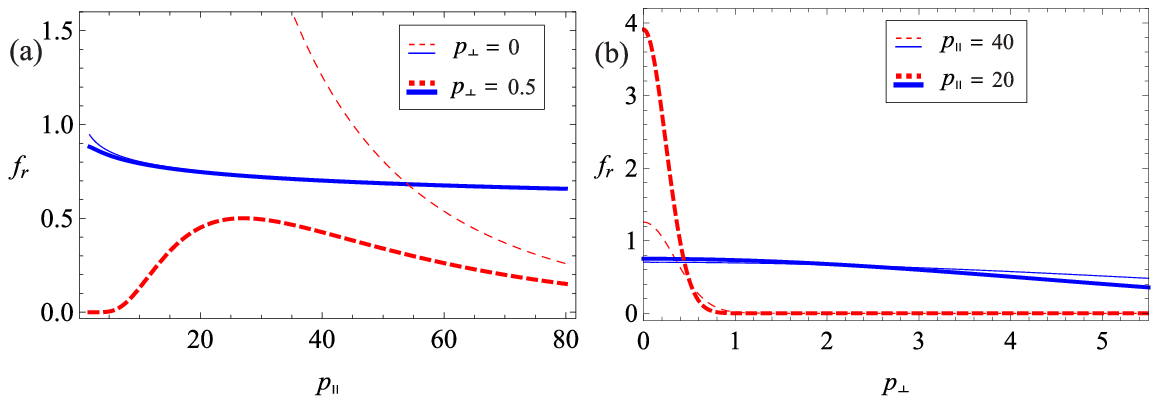}
\caption{Comparison between the near-critical, $f_r/A$ (blue solid) and
  avalanche, $100 f_r^{\mathrm{disr}}$ (red dashed) distribution functions. For the near-critical
  distribution we used $Z=1$ and $\alpha=1.3$. For the
  avalanche distribution we used $\ln{\Lambda}=18$, $Z=1$ and
  $E=40\;\rm V/m$ (corresponding to $\alpha=865$). (a) The
  distribution function as a function of $p_\parallel$ for $p_\perp=0$
  (thin lines) and $p_\perp=0.5$ (thick lines). (b) The distribution
  function as a function of $p_\perp$ for $p_\parallel=20$ (thick lines) and $p_\parallel=40$ (thin lines).}
\label{fig:nearandaval}
\end{center}
\end{figure}

Comparing the two distributions, we note that the near-critical
distribution function in Equation~(\ref{eq:near-distribution})
represents a broader beam, with a less rapidly decaying tail.
Figures~\ref{fig:distr_critfield}-\ref{fig:nearandaval} show the
comparison between the near-critical and the avalanching distribution
functions.  Figure \ref{fig:distr_critfield}a shows the near-critical
distribution for two different values of $\alpha$ and $Z=1.5$. Figure
\ref{fig:distr_critfield}b shows the comparison between the
$\alpha=1.3$, $Z=1$ case {of the near-critical distribution} with the
avalanching distribution, which is significantly more beamlike. To
illustrate that the distribution is more beamlike and more rapidly
decaying in $p_\parallel$ in the avalanche case, in
Figure~\ref{fig:nearandaval} we show the comparison between the
near-critical and avalanching distributions for specific values of
$p_\parallel$ and $p_\perp$. In spite of the differences noted above,
the distribution functions in the $\alpha\mygtrsim 1$ and $\alpha\gg
1$ limits are similar in the sense that both have an anisotropy in the
$p_\parallel$ direction, and a smooth transition between the two can
be envisioned based on Figure~\ref{fig:distr_critfield}. The reason
for using this particular distribution function (Eq.~(\ref{eq:near-distribution})) in the
present work is that it is at the lower limit of $\alpha$ that can
possibly produce runaway electrons.

Although Equation~(\ref{eq:near-distribution}) describes
  primary generation of runaways, it does not mean that the generation
  rate is small. Primary generation implies that the runaway
  generation is smaller than $n_e/\tau$. But since $n_e/\tau$ is very
  large, primary generation can result in a substantial runaway
  electron population and its importance has been shown in many
  numerical simulations, see e.g. \cite{hakan}.
\subsection{Resonance condition}
In a plasma with a slightly supercritical electric field, the
characteristic value of the normalized momentum $p$ in the runaway
region satisfies $p>1/\sqrt{\alpha-1}$. To obtain an explicit formula
for the resonant momentum, the expression $\gamma = \sqrt{1 + p_\perp
  ^2 + p_\parallel ^2}$ should be substituted into the resonance
condition, and  that leads to
\begin{equation} \label{eq:pres}
p_{\mathrm{res}} \left( p_{\perp}, k_{\parallel}, \omega _0 \right) = \frac{- k_{\parallel} c m \omega _{ce} \pm \omega _0 \sqrt{(k_{\parallel} ^2 c^2 - \omega _0 ^2) (1 + p_{\perp} ^2) + m^2 \omega _{ce} ^2}}{k_{\parallel} ^2 c^2 - \omega _0 ^2}.
\end{equation}
By using this general resonance condition, the expressions giving the
imaginary parts of the runaway susceptibilities become quite
complicated. The full expressions for the susceptibilities are given
in Appendix A.

Only the $p_{\mathrm{res}} > 0$ resonant momenta are physically relevant. By
studying the $p_{\mathrm{res}} > 0$ condition for different signs of $m$, using
the relation between $k_{\parallel} c$ and $\omega _0 (k, \theta)$ it
can be shown that the Doppler resonances ($m>0$) cannot be satisfied
for any of the solutions in Equation~(\ref{fomega}) or Equation~(\ref{momega}).

\paragraph{Anomalous Doppler resonance}
For the anomalous Doppler resonance ($m<0$) the $p_{\mathrm{res}} > 0$
condition, defining the physically relevant region of the $p_{\mathrm{res}}$
resonant momentum, is
\begin{equation}
\frac{k_{\parallel} c |m| \omega _{ce} + \omega _0 \sqrt{(k_{\parallel} ^2 c^2 - \omega _0 ^2) (1 + p_{\perp} ^2) + m^2 \omega _{ce} ^2}}{k_{\parallel} ^2 c^2 - \omega _0 ^2} > 0,
\end{equation}
leading to $k_{\parallel}^2 c^2 > \omega _0^2 (k, \theta)$, which is
only satisfied for the electron-whistler branch and not for the other two
solutions of Equation~(\ref{fomega}). Also the magnetosonic-whistler wave
can be destabilized via this resonance.

\paragraph{Cherenkov resonance}

For Cherenkov resonance (the case of $m=0$) the $p_{\mathrm{res}} > 0$
condition is
\begin{equation}
\frac{\omega _0 \sqrt{(k_{\parallel} ^2 c^2 - \omega _0 ^2) (1 + p_{\perp} ^2)}}{k_{\parallel} ^2 c^2 - \omega _0 ^2} > 0.
\end{equation}
This also leads to the condition $k_{\parallel}^2 c^2 > \omega _0^2
(k, \theta)$, narrowing down the possible waves once again to the
electron-whistler wave and the magnetosonic-whistler wave.
Summarizing the results above, we conclude that for $m \leq 0$, only
the electron-whistler waves can yield physically relevant results out
of the high frequency electron waves defined by the dispersion
relation in Equation~(\ref{fomega}). The magnetosonic-whistler waves can
also be destabilized via $m \leq 0$ resonances.
However, combining the region of the validity of the wave frequency
with the resonance condition it can be seen that in the
magnetosonic-whistler case the destabilization is most effective by
very energetic (around $10 \;\rm MeV$) runaway electrons.

If $p\gg 1$, for the beam-like distribution function in
Equation~(\ref{eq:near-distribution}) with $p_\parallel\gg p_\perp$, the
$\gamma\sim \left|p_\parallel\right|$ approximation can be used (which
will be called the ultra-relativistic limit), and the resonance
condition in (\ref{eq:runaway-susceptibility}) simplifies to
\begin{equation}
p_\parallel=\frac{-m\omega_{ce}}{k_\parallel c-\omega}
\label{eq:resonance}
\end{equation}
and  $m<0$ for physically relevant results. 
\section{Unstable waves}
\label{sec:growth}
\begin{figure}[htbp]
\begin {center}
  \includegraphics[width=0.9\textwidth]{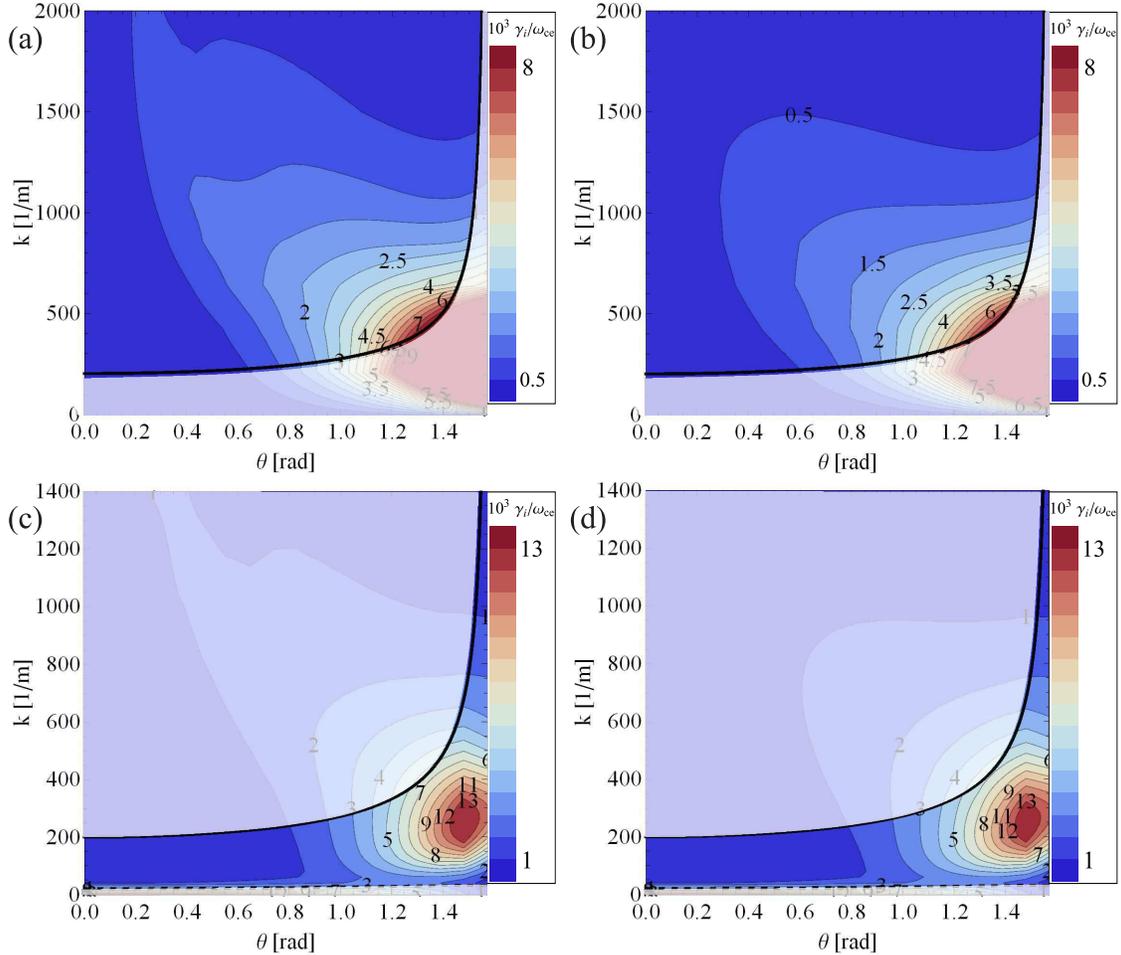}
\caption{Normalized growth rate $10^3 \gamma _i / \omega _{ce}$ for the
  electron-whistler wave (a,b) and the magnetosonic-whistler wave
  (c,d). Both in (a,b) and (c,d) the black line is $\omega=\omega
  _{ce} / 45$, the electron-whistler approximation is valid in the
  region above it.  In (c,d) the dashed line denotes $\omega= \omega
  _{ci}$, the magnetosonic-whistler approximation is valid in the
  region above it. 
The rest of the
  parameters are $n_e = 5 \cdot 10^{19}$ ${\rm m}^{-3}$, $n_r = 3
  \cdot 10^{17}$ ${\rm m}^{-3}$, $B = 2$ T and $p_{\mathrm{max}}=5$. (a,c) Ultrarelativistic
  resonance condition for $m = -1$. (b,d) General resonance condition,
  sum of the cases $m = -1$ and $m = 0$.}
\label{fig: electron_wh_nearcrit}
\end {center}
\end{figure}
The instability growth rates for the electron-whistler and the
magnetosonic-whistler waves can be calculated from
Eqs.~(\ref{eq:growth-electron}) and (\ref{eq: growth-magneto}) as
functions of $\textbf{k}$.  Figure \ref{fig: electron_wh_nearcrit}ab
shows the growth rates for the electron-whistler wave using the
ultrarelativistic limit and the general resonance condition,
respectively, for the baseline parameters. The growth
rate increases with decreasing $k$ throughout the range of validity of
the electron-whistler approximation.  Comparing Figure~\ref{fig:
  electron_wh_nearcrit}a and Figure~\ref{fig: electron_wh_nearcrit}b it
can be seen that by using the ultrarelativistic condition one gets
somewhat different results than by using the general condition, but
the qualitative behaviour is the same.

Figure \ref{fig: electron_wh_nearcrit}cd shows the growth rates for
magnetosonic-whistler wave. In contrast to the electron-whistler wave,
the growth rate has a well-defined maximum. However, as it was
mentioned before, the resonance condition cannot be satisfied in the
region close to the maximum growth rate unless the resonant energy is
very high (around 10 MeV for the parameters given in the figure
caption), which is only expected to be reached by few electrons in the
near-critical case considered in this paper. Interestingly in both
cases, the largest instability growth rate occurs for the region in
the $k$-$\theta$-space where the whistler approximation (\ref{ww}) is
valid (see the low $k$ and perpendicular propagation part of
Figure~\ref{fig:whistlers}). However, we note that only high energy
electrons can interact with the quasi-perpendicularly propagating
whistler wave, therefore it is more likely that the electron-whistler
branch with slightly higher $k$ and more oblique propagation is the
one which is the most unstable wave.

\subsection{Most unstable wave}
\begin{figure}[htbp]
\begin{center}
\includegraphics[width=0.45\textwidth]{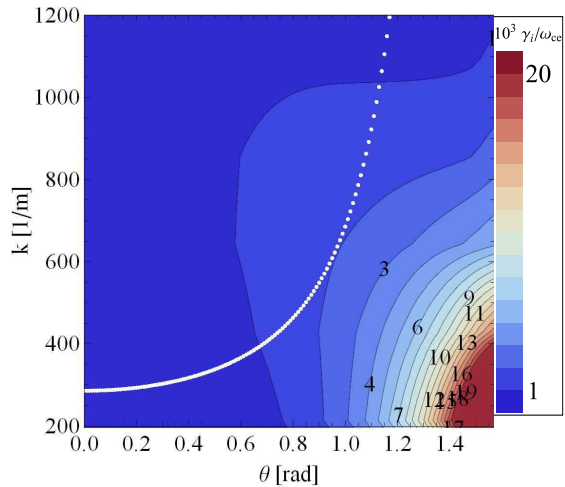}
\caption{Most unstable wave in the near-critical case: maximum of the growth rate ($10^3 \gamma _i / \omega _{ce}$, contour lines) on the line corresponding to the maximum runaway energy ($2.6$ MeV, white dots). The parameters are $n_e = 5 \cdot 10^{19}$ ${\rm m}^{-3}$, $n_r = 3 \cdot 10^{17}$ ${\rm m}^{-3}$, $B = 2$ T.}
\label{fig: most_unstable}
\end{center}
\end{figure}

The normalized momentum corresponding to the maximum energy of $2.6$ MeV
is approximately $p_{\mathrm{res}} = 5$. The corresponding wave numbers and
propagation angles of the electron-whistler wave can be calculated by
using the general resonance condition and the dispersion relation. The
growth rate and the values corresponding to $p_{\mathrm{res}} = 5$ are shown
on Figure~\ref{fig: most_unstable}. The most unstable wave in the
near-critical case is an electron-whistler wave with frequency $4.2
\cdot 10^{10}$ ${\rm s}^{-1} \simeq 0.12 \omega_{ce}$ (for a magnetic
field of $2\;\rm T$), wave number of approximately $650$
${\rm m}^{-1}$, and angle of propagation $\theta \approx 0.9$.

It should be noted, that the parameters of the most unstable wave are
sensitive to the magnetic field. The reason is that the resonance
condition is highly dependent on the magnetic field through the
gyrofrequency. Due to this fact, the $p < 5$ condition for the
momentum of the runaway electrons yields very different wave numbers
for the most unstable wave. For example, if the magnetic field is $4$
T instead of the $2$ T on Figure \ref{fig: most_unstable}, the $p_{\mathrm{res}}
= 5$ resonant momentum yields $k \sim 1600$ ${\rm m}^{-1}$ wave
number and $\theta \sim 0.3$. Therefore, by increasing the magnetic field, the
wave number and frequency of the most unstable wave increase,
while the angle of propagation decreases.

\begin{figure}[htbp]
\begin{center}
\includegraphics[width=0.95\textwidth]{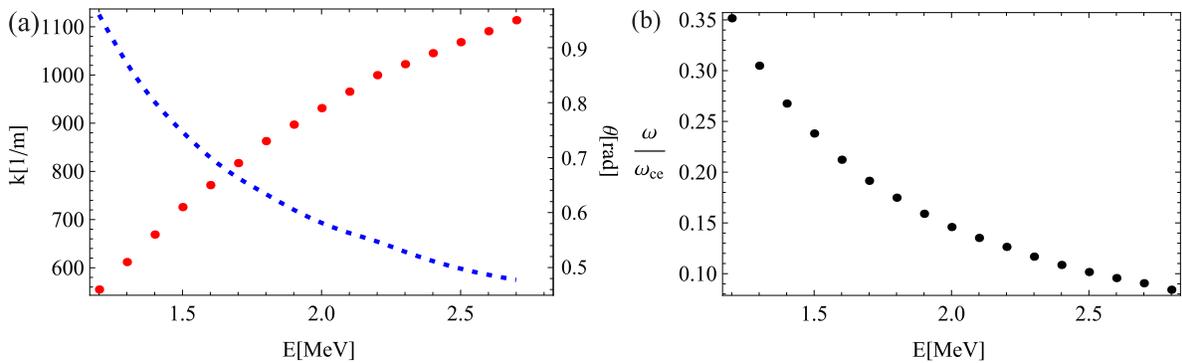}
\caption{The value of wave number (blue dashed) and propagation angle (red dotted) (a) and
  frequency (b) of the most unstable wave as function of maximum
  runaway energy.}
\label{fig:stability1}
\end{center}
\end{figure}

The wave number, propagation angle and frequency of the most unstable
wave also depend on the maximum runaway energy. Figure
\ref{fig:stability1} shows that as the energy grows the propagation
angle becomes larger and the wave number and frequency drops. This
means that for low energy runaway electrons (energies just above the
critical energy for runaway acceleration) we expect frequencies
slightly less than half of the electron cyclotron frequency,
propagating angles of $\theta \simeq 0.4$ and wave numbers of 1000
$\rm m^{-1}$.  As the runaway energy grows the frequency and wave
number of the most unstable wave falls.

\subsection{Stability diagram}

In order to determine the stability limits, the instability growth
rate of the wave has to be compared to the damping rates.  In cold
plasmas, collisional damping is dominant, and the damping is
approximately equal to $ \gamma_d=1.5 \tau_{ei}^{-1}$
\cite{brambilla}, where
$\tau_{ei}=3\pi^{3/2}m_{e0}^{2}v_{Te}^{3}\epsilon_{0}^2/n_iZ^2e^4\ln{\Lambda}$
is the electron-ion collision time.  In addition to collisional
damping, the wave is damped due to the fact that the extent of the
runaway beam is finite, and the wave energy is transported out of its
region with a $\partial \omega/\partial k_\perp$ perpendicular group
velocity. This mechanism can be accounted for by adding a convective
damping term $\gamma_v\equiv(\partial \omega/\partial k_\perp)/(4
L_r)$, where $L_r$ is the radius of the runaway beam \cite{fulop1}.
The linear growth rate of a wave is thus $
\gamma_{l}=\gamma_{i}-\gamma_{d}-\gamma_{v}$, and the wave is
unstable, if $\gamma_{l}>0$.  A simple estimate of the order of
magnitude of the damping rates shows that for typical parameters, and
for reasonably narrow electron beams, the convective damping is
expected to dominate if $T_e>200 \;\rm eV$. However, as the plasma
temperature seldom reaches such a high value in the relevant case of
tokamak disruptions \cite{ward}, collisional damping should not be
neglected.

\begin{figure}[htbp]
\begin{center}
\includegraphics[width=0.9\textwidth]{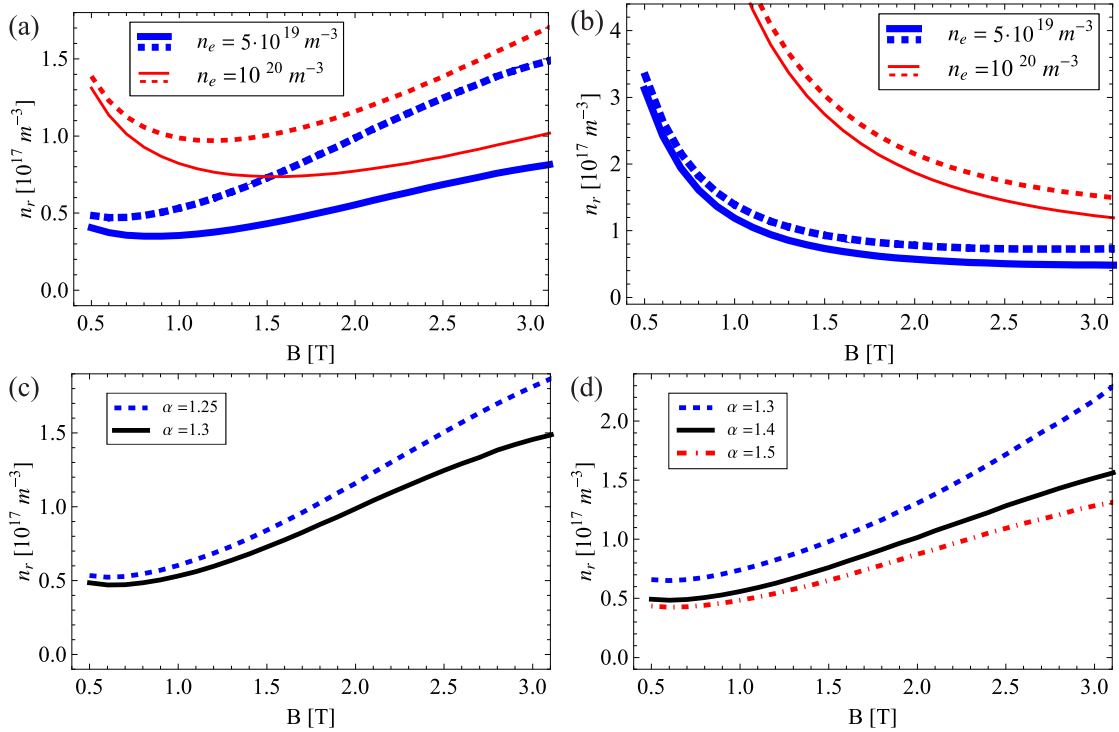}
\caption{Stability thresholds for the most unstable wave in
  near-critical electric field, for electron temperature $T_e=20\;\rm eV$. (a,b) Stability threshold as function
  of magnetic field for the electron-whistler wave and
  magnetosonic-whistler waves, respectively. The runaway-beam radius
  is $L_r = 0.1$ m (dashed) and $L_r = 0.2$ m (solid). In (a) we
  assume $p_{\mathrm{max}}=5$ and in (b) $p_{\mathrm{max}}=20$. (c,d) Sensitivity of the
  stability threshold to the normalized electric field $\alpha$ for the electron-whistler wave. The
  runaway beam radius is $L_r=0.1 \;\rm m$, $n_e=5\cdot 10^{19}\;\rm
  m^{-3}$ and the maximum runaway energy is $2.6\;\rm MeV$, corresponding to $p_{\mathrm{max}}=5$. In (c) $Z=1$ and in (d) $Z=1.5$. }
\label{fig: stability}
\end{center}
\end{figure}

The linear stability threshold was determined in the following
way. For any given value of the magnetic field the growth rate is
calculated (for $m = -1$, $m = 0$,
then adding them for all $k$ and $\theta$), then the collisional and
convective damping rates are subtracted from it. The parameters of the
most unstable wave are then determined. The stability threshold of the
most unstable electron-whistler wave is shown on Figure~\ref{fig:
  stability}a. We conclude that for typical parameters the runaway
density needed to counter the damping rates, therefore to destabilize
an electron-whistler wave is of the order of $10^{17}$ ${\rm m}^{-3}$
or $n_r/n_e=0.2 \%$. For the magnetosonic-whistler wave, the stability
threshold is shown in Figure~\ref{fig: stability}b. In this case we
assumed $p_{\mathrm{max}}=20$, since in this case the resonant
particles have higher energies than in the electron-whistler
case. Figure \ref{fig: stability}cd shows the dependence of the
stability threshold on the normalized electric field $\alpha$. We
conclude that above $B\mygtrsim 1.5$, the runaway density needed for
destabilization is sensitive to $\alpha$, for higher normalized
electric field lower runaway density is needed to destabilize the
wave. This dependence on $\alpha$ is due to the fact
  that for a higher electric field the anisotropy of the runaway
  distribution and thus the destabilizing effect is stronger,
  therefore a lower density of runaways suffices for a resonant
  destabilization of the wave. This is by no means because of the
  special characteristics of the model distribution we used, but is
  due to the underlying physics.
	
\begin{figure}[htbp]
\begin{center}
\includegraphics[width=0.45\textwidth]{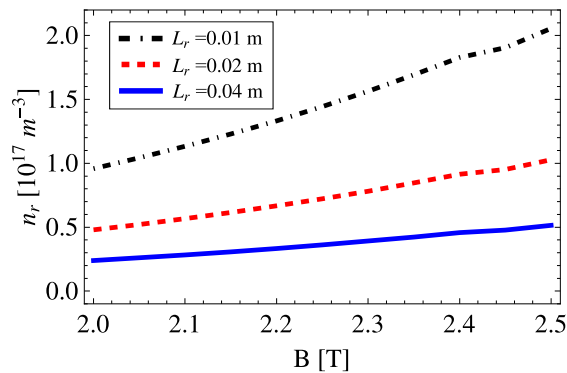}
\caption{Stability threshold for the most unstable electron-whistler wave in
  near-critical electric field, for the experimental parameters of the T-10 tokamak. The parameters are $\alpha=1.9$, $Z=3$, $n_e = 4 \cdot 10^{19} \;\rm m^{-3}$, $T_e=0.5\;\rm keV$, $p_{\mathrm{max}}=1.5$.}
\label{fig: stability_T10}
\end{center}
\end{figure}

  It is instructive to compare the order of magnitude
    of the runaway density required for destabilization of the
    whistler wave to the one that is measured in an experimental
    setup. Figure~\ref{fig: stability_T10} shows the stability
    thresholds as function of magnetic field, for various runaway beam
    radii, for parameters relevant to an experiment in the T-10
    tokamak \cite{savrukhin}.  We note that the runaway density needed
    for destabilization is about $10^{17} \;\rm m^{-3}$ even for a
    narrow runaway beam. The runaway density estimated in the
    experiment was almost an order of magnitude higher than this: $7
    \cdot 10^{17} \;\rm m^{-3}$.

\section{Conclusions}
\label{sec:conclusions}
The presence of high energy electrons is often associated with bursts
of high-frequency waves. The emission of radiation is most often due
to Bremsstrahlung and synchrotron radiation, but in certain cases they
are due to instabilities caused by the anisotropy. The observation of
these waves can help to determine the origin and evolution of the
energetic electrons, and also in some cases, the properties of the
background plasma \cite{ww}. The instability may
  result in pitch-angle scattering induced isotropization and may
  therefore prevent the harmful effects of the runaway electron beam
  \cite{pokol}.

The reason for the generation of an anisotropic runaway electron
population is the high electric field {that is often caused by
  reconnection events in magnetized plasmas}. In previous calculations
regarding waves driven by runaways the electric field was assumed to
be much higher than the critical field $\alpha\gg 1$. This is not
often the case in reality. Therefore, in this paper we use an
electron distribution function that is valid in the near-critical
case. We show that in this case, the distribution is broader and less
rapidly decaying compared to the $\alpha\gg 1$ case. 

By studying the linear growth rate of the electron-whistler branch
(valid in the frequency region $\omega_{ce} {\sqrt{m_e/m_i}} \ll
\omega $) and the magnetosonic-whistler branch (valid in the frequency
region $\omega_{ci}\ll \omega\ll\omega_{ce}$) separately, we find that
the frequency of the most unstable wave is in the region where these
overlap and have characteristics similar to the whistler
approximation. For typical tokamak parameters we find that the
frequency of the most unstable wave is around $0.12 \omega_{ce}$, in
the near-critical case with $\alpha=1.3$ and $E=2.6\;\rm MeV$. The
frequency and wave number of the most unstable wave depends strongly
on the magnetic field and on the maximum runaway energy. By comparing
the ultra-relativistic limit of the resonance condition and the
general one we show that although the behaviour of the instability
growth rates of the electron-whistler and magnetosonic-whistler waves
are similar, the actual values for the growth rate may differ, and
therefore the frequency and wave number of the most unstable wave
might be different.

The instability growth rate of the electron-whistler wave was compared
to the collisional and convective damping rates. We find that the
number density of runaways that is required to destabilize the waves
increases with increasing magnetic field. For low magnetic fields the
convective damping decreases, while the collisional damping rate
remains constant, making it dominant in this region.  As the growth
rate also decreases, the stability limit is high for low magnetic
fields. We investigated the stability of the whistler
  waves for parameters relevant to the T-10 tokamak \cite{savrukhin},
  where the effective electric field is near-critical. We found that
  the observed runaway density is about an order of magnitude higher
  than the density needed for the most unstable electron-whistler wave
  to be destabilized. Thus the runaway population may indeed give rise
  to this whistler wave.

The importance of this study is that it considers the case where the
electric field is near critical, which is opposite to the other limit
that has been considered in previous work \cite{pokol,fulop,fulop1}
(when the electric field is far above the critical). By investigating
this case, we show that the high-frequency instabilities are
qualitatively similar, but have different frequencies and wave
numbers. This result may open up the possibility of
diagnostics. Understanding the properties of the waves destabilized by
runaway electrons can be important in view of obtaining information
about the energetic electron population and the background
plasma. Regardless of the fact that the distribution used in this work
is only valid in the near-critical case, if we compare it to the
avalanche distribution we observe a smooth transition between the two,
and so we expect that the distribution does not change
qualitatively. Also, the characteristics of the growth rates in the
near-critical and high electric field limits are similar, in the sense
that the maximum of the growth rate is at low wave numbers and
near-perpendicular propagation in both cases \cite{fulop}. The
differences in the parameters of the most unstable wave for
near-critical and avalanching cases are mainly due to the maximum
runaway energy. The similarity of the results, added to the relaxation
of the approximations used in previous work, opens the way toward more
general numerical studies of wave-particle interaction for arbitrary
electric fields. 

The whistler waves, if they grow to significant amplitude, in
principle could perturb the background magnetic field and lead to
efficient transport of particles (specially energetic ones) out from
the plasma. The effect of magnetic field perturbation has been studied
before \cite{radialdiffusion,papp1,papp2}, and it has been shown that
runaway avalanches can be prevented altogether with sufficiently
strong radial diffusion. However, the magnetic fluctuation level that
is required for this to happen is estimated to be $\delta B/B\sim
10^{-3}$ and the magnetic fluctuation level induced by these
high-frequency whistler waves would be several orders of magnitude
lower than this value. Therefore, as mentioned before, the runaway
electron population will be affected mostly through pitch-angle
scattering and concomitant isotropization and synchrotron radiation
damping.

\section*{Acknowledgements}
This work, supported by the European Communities under the contract of
association between EURATOM, {\em Vetenskapsr{\aa}det} and the
Hungarian Academy of Sciences, was carried out within the framework of
the European Fusion Development Agreement. The authors
  are grateful to H. Smith, G. Papp and P. Helander for fruitful
  discussions. One of the authors acknowledges the financial support
from the FUSENET Association. The views and opinions expressed herein
do not necessarily reflect those of the European Commission.

\section*{Appendix A: Runaway susceptibilities}
\label{sec:appa}

In the general case the susceptibilities have the following form:

\begin{eqnarray}
\hspace{-1cm}  {\rm Im} \chi _{11} ^r \left( \textbf{k}, \omega _0 \right) &=& - \frac{2 \pi ^2 \omega _{pr} ^2 \omega _{ce} ^2}{\omega _0 ^2 k_{\perp} ^2 c^2} \int ^{\infty} _0 dp_{\perp} \int ^{\infty} _{- \infty} dp_{\parallel} \sum m^2 J_m ^2 (z) \times \\
  \nonumber
 & & \left( \frac{\partial f_r (\textbf{p})}{\partial p_{\perp}} \left( \frac{m \omega _{ce}}{\gamma} \right) + \frac{\partial f_r (\textbf{p})}{\partial p_{\parallel}} \frac{k_{\parallel} c p_{\perp}}{\gamma} \right) \frac{1}{\gamma} \cdot \delta \left( \omega _0 - \frac{k_{\parallel} c p_{\parallel}}{\gamma} - \frac{m \omega _{ce}}{\gamma} \right)
\end{eqnarray}
where we used the resonance condition $\gamma \omega _0 -
k_{\parallel} c p_{\parallel} = m \omega _{ce}$ to replace the factor
$ \left( \omega _0 - k_{\parallel} c p_{\parallel}/\gamma\right)$.
The ${\rm Im} \chi _{22} ^r$ and ${\rm Im} \chi _{12} ^r$ terms
only differ in multiplicative constants and will be presented later.

For a general function $g(p_\parallel)$ we can rewrite the integral in $p_\parallel$ as follows
\begin{equation} \label{eq:susc_integral} \int dp_{\parallel} \delta
  \left( \omega _0 - \frac{k_{\parallel} c p_{\parallel}}{\gamma} -
    \frac{m \omega _{ce}}{\gamma} \right) g(p_{\parallel}) = \int dx
  \delta \left( A - \frac{B x}{\sqrt{C+x^2}} + \frac{D}{\sqrt{C+x^2}}
  \right) g(x).
\end{equation}
where $x = p_{\parallel}$, $A = \omega _0$, $B = k_{\parallel} c$, $C
= 1 + p_{\perp} ^2$ and $D = - m \omega _{ce}$.
Changing variables
$
y = (B x - D)/\sqrt{C + x^2},$ so that
\begin{eqnarray}
\nonumber
dx &=& \frac{(C+ x^2)^{3/2}}{B C + x D} dy, \\
\nonumber
x &=& \frac{B D \pm y \sqrt{(B^2 - y^2) C + D^2}}{B^2 - y^2},
\end{eqnarray}
Equation (\ref{eq:susc_integral}) yields
\begin{eqnarray}
& & \int dy \, \frac{(C + x^2)^{3/2}}{B C + x D} \; \delta (A - y) g\left( \frac{B D \pm y \sqrt{(B^2 - y^2) C + D^2}}{B^2 - y^2} \right) = \\
\nonumber
& &  \frac{\left( A D \pm B \sqrt{(B^2 - A^2) C + D^2} \right)^3}{(B^2 - A^2)^3 \cdot \left[ B C + \frac{D}{B^2 - A^2} \left( B D \pm A \sqrt{(B^2 - A^2) C + D^2} \right) \right]} g\left( \frac{B D \pm A \sqrt{(B^2 - A^2) C + D^2}}{B^2 - A^2} \right).
\end{eqnarray}
Using the above expression to solve the integrals in the runaway
susceptibilities, we arrive to the following formulas:
\begin{eqnarray} \label{eq: susc_general11}
{\rm Im} \chi _{11} ^r \left( \textbf{k}, \omega _0 \right) &=& - \frac{2 \pi ^2 \omega _{pr} ^2 \omega _{ce} ^2}{\omega _0 ^2 k_{\perp} ^2 c^2} \int ^{\infty} _0 dp_{\perp} \sum m^2 J_m ^2 (z) \times \\
\nonumber
& & \left[ \left( \frac{\partial f_r (\textbf{p})}{\partial p_{\perp}} \left( \frac{m \omega _{ce}}{\gamma} \right) + \frac{\partial f_r (\textbf{p})}{\partial p_{\parallel}} \frac{k_{\parallel} c p_{\perp}}{\gamma} \right) \right]_{p_{\parallel} = p_{\mathrm{res}}}  \frac{h\left(p_{\perp}, k_{\parallel}, \omega _0 \right)}{\gamma},
\end{eqnarray}
where
\begin{eqnarray}
h\left(p_{\perp}, k_{\parallel}, \omega _0 \right) &=& \frac{1}{(k_{\parallel} ^2 c^2 - \omega _0 ^2)^3} \times \\
\nonumber
& & \hspace{-2cm}  \frac{\left( - \omega _0 \, m \omega _{ce} \pm k_{\parallel} c \sqrt{\left( k_{\parallel} ^2 c^2 - \omega _0 ^2 \right) (1 + p_{\perp} ^2) + m^2 \omega _{ce} ^2} \right)^3}{\left[ k_{\parallel} c \, (1 + p_{\perp} ^2) - \frac{m \omega _{ce}}{k_{\parallel} ^2 c^2 - \omega _0 ^2} \left( - k_{\parallel} c \, m \omega _{ce} \pm \omega _0 \sqrt{\left( k_{\parallel} ^2 c^2 - \omega _0 ^2 \right) (1 + p_{\perp} ^2) + m^2 \omega _{ce} ^2} \right) \right]}.
\end{eqnarray}
Similarly the other two terms of the runaway susceptibility:

\begin{eqnarray} \label{eq: susc_general22}
{\rm Im} \chi _{22} ^r \left( \textbf{k}, \omega _0 \right) &=& - \frac{2 \pi ^2 \omega _{pr} ^2}{\omega _0 ^2} \int ^{\infty} _0 p_{\perp} ^2 dp_{\perp} \sum (J' _m (z))^2 \times \\
\nonumber
& & \left[ \left( \frac{\partial f_r (\textbf{p})}{\partial p_{\perp}} \left( \frac{m \omega _{ce}}{\gamma} \right) + \frac{\partial f_r (\textbf{p})}{\partial p_{\parallel}} \frac{k_{\parallel} c p_{\perp}}{\gamma} \right) \right]_{p_{\parallel} = p_{\mathrm{res}}}  \frac{h\left(p_{\perp}, k_{\parallel}, \omega _0 \right)}{\gamma},
\end{eqnarray}

\begin{eqnarray} \label{eq: susc_general12}
- {\rm Re} \chi _{12} ^r \left( \textbf{k}, \omega _0 \right) &=& - \frac{2 \pi ^2 \omega _{pr} ^2 \omega _{ce}}{\omega _0 ^2 k_{\perp} c} \int ^{\infty} _0 p_{\perp} dp_{\perp} \sum m J_m (z) J' _m (z) \times \\
\nonumber
& & \left[ \left( \frac{\partial f_r (\textbf{p})}{\partial p_{\perp}} \left( \frac{m \omega _{ce}}{\gamma} \right) + \frac{\partial f_r (\textbf{p})}{\partial p_{\parallel}} \frac{k_{\parallel} c p_{\perp}}{\gamma} \right) \right]_{p_{\parallel} = p_{\mathrm{res}}}  \frac{h\left(p_{\perp}, k_{\parallel}, \omega _0 \right)}{\gamma}.
\end{eqnarray}
Equation (\ref{eq: susc_general12}) gives the real part of the runaway
susceptibility $\chi_{12}$ since this term is the one needed in the
expression for the growth rate.  By substituting these runaway
susceptibilities into the expression of the growth rate, the
calculations yield results in the general, relativistic case regarding
the runaway electrons interacting with the corresponding wave.


\begin{thebibliography}{11}
\bibitem{td} J.A. Wesson, R.D. Gill, M. Hugon, F.C. Schüller, J.A. Snipes, D.J. Ward, D.V. Bartlett, D.J. Campbell, P.A. Duperrex, A.W. Edwards, R.S. Granetz. N.A.O. Gottardi, T.C. Hender, E. Lazzaro, P.J. Lomas, N. Lopes Cardozo, K.F. Mast, M.F.F. Nave, N.A. Salmon, P. Smeulders, P.R. Thomas, B.J.D. Tubbing, M.F. Turner and A. Weller, {\it Nucl. Fusion} {\bf 29}, 641 (1989).
\bibitem{tavani} M. Tavani, M. Marisaldi, C. Labanti, F. Fuschino, A. Argan, A. Trois, P. Giommi, S. Colafrancesco, C. Pittori,
F. Palma, M. Trifoglio, F. Gianotti, A. Bulgarelli, V. Vittorini, F. Verrecchia, L. Salotti, G. Barbiellini,
P. Caraveo, P. W. Cattaneo, A. Chen, T. Contessi, E. Costa, F. D’Ammando, E. Del Monte, G. De Paris,
G. Di Cocco, G. Di Persio, I. Donnarumma, Y. Evangelista, M. Feroci, A. Ferrari, M. Galli, A. Giuliani,
M. Giusti, I. Lapshov, F. Lazzarotto, P. Lipari, F. Longo, S. Mereghetti, E. Morelli, E. Moretti, A. Morselli, L. Pacciani, A. Pellizzoni, F. Perotti, G. Piano, P. Picozza, M. Pilia, G. Pucella, M. Prest,
M. Rapisarda, A. Rappoldi, E. Rossi, A. Rubini, S. Sabatini, E. Scalise, P. Soffitta, E. Striani, E. Vallazza,
S. Vercellone, A. Zambra and D. Zanello, {\it Phys. Rev. Lett.} {\bf 106}, 018501 (2011).
\bibitem{moghaddam} E. Moghaddam-Taaheri and C.K. Goertz, {\it Astrophys. J.} {\bf 352}, 361 (1990).
\bibitem{pokol} G. Pokol, T. F\"ul\"op and M. Lisak, {\it Plasma Phys.~Control.~Fusion} {\bf 50}, 045003 (2008).
\bibitem{fulop} T. F\"ul\"op, G. Pokol, P. Helander and M. Lisak, {\it Phys.~Plasmas} {\bf 13},  062506 (2006).
\bibitem{fulop1} T. F\"ul\"op, H.M. Smith and G. Pokol, {\it Phys.~Plasmas} {\bf 16},  022502 (2009). 
\bibitem{gill} R.D. Gill, B. Alper, M. de Baar, T.C. Hender, M.F. Johnson,
V. Riccardo and contributors to the EFDA-JET Workprogramme, {\it Nucl. Fusion} {\bf 42}, 1039 (2002).
\bibitem{jt60} R. Yoshino, S. Tokuda and Y. Kawano, {\it Nucl. Fusion} {\bf 39},
  151 (1999).
\bibitem{savrukhin} P.V. Savrukhin, {\it Phys. Rev. Lett.} {\bf 86}, 14 (2001).
\bibitem{riemann} J. Riemann, H.M. Smith and P. Helander, {\it Phys.~Plasmas} {\bf 19}, 012507 (2012).
\bibitem{sandquist} P Sandquist, S.E. Sharapov, P. Helander and M. Lisak, {\it Phys.~Plasmas} {\bf 13} 072108 (2006).
\bibitem{stix} T. H. Stix, {\it Waves in plasmas}, American Institute of
  Physics, New York, 1992.
\bibitem{ww} S. Sazhin, {\it Whistler-mode waves in a hot plasma}, Cambridge University Press, Cambridge, 1993.
\bibitem{connorhastie} W. Connor and R.J. Hastie, {\it Nucl. Fusion} {\bf
    15}, 415 (1975).
\bibitem{helander} P. Helander, L.-G. Eriksson and F. Andersson, {\it Plasma Phys.~Control.~Fusion}
  {\bf 44}, B247 (2002).
\bibitem{rosput} M.N. Rosenbluth and S.V. Putvinski, {\it Nucl. Fusion} {\bf 37}, 1355 (1997).
\bibitem{hakan} H.M. Smith, P. Helander, L.-G. Eriksson, D. Anderson, M. Lisak and F. Andersson, {\it Phys. Plasmas} {\bf 13} 102502 (2006).
\bibitem{brambilla} M. Brambilla, {\it Phys. Plasmas} {\bf 2}, 1094 (1995).
\bibitem{ward} D.J. Ward and J.A. Wesson, {\it Nucl. Fusion} {\bf 32} 1117 (1992).
\bibitem{radialdiffusion} P. Helander, L.-G. Eriksson and F. Andersson, {\it Phys. Plasmas} {\bf 7}, 4106 (2000).
\bibitem{papp1}G. Papp, M. Drevlak, T. F\"ul\"op, P. Helander and G.I. Pokol, {\it Plasma Phys.~Control.~Fusion} {\bf 53}, 095004 (2011).
\bibitem{papp2}G. Papp, M. Drevlak, T. F\"ul\"op and P. Helander, {\it Nucl. Fusion} {\bf 51} 043004 (2011).
\end{thebibliography}
\end{document}